\documentclass[journal]{IEEEtran}

\usepackage{pgfplots,pgfplotstable,booktabs}

\pgfplotsset{compat=1.5}

\usepackage{amsmath, amssymb}

\usepackage{caption}

\usepackage{graphicx}

\usepackage{subfigure}

\usepackage{graphicx,epstopdf}

\DeclareGraphicsExtensions{.eps}

\usepackage{epsf}  

\graphicspath{{images/}}

\usepackage{etoolbox}

\apptocmd{\sloppy}{\hbadness 10000\relax}{}{}

\apptocmd{\sloppy}{\hbadness 6236\relax}{}{}

\usepackage{cite}

\begin{document}

\title{Numerical Polynomial Homotopy Continuation Method to Locate All The Power Flow Solutions}

\author{Dhagash~Mehta, Hung~D.~Nguyen,~\IEEEmembership{Student Member,~IEEE,} and~Konstantin~Turitsyn,~\IEEEmembership{Member,~IEEE}
\thanks{Dhagash Mehta is with the Department of Applied and Computational Mathematics and Statistics, University of Notre Dame, Notre Dame, IN 46556, USA, and with the Department of Chemistry, The University of Cambridge, Lensfield Road, Cambridge CB2 1EW, UK. email: Dhagash.B.Mehta.11@nd.edu.

Hung D. Nguyen and Konstantin Turitsyn are with the Department of Mechanical Engineering, Massachusetts Institute of Technology, Cambridge, MA, 02139 USA e-mail: hunghtd@mit.edu and turitsyn@mit.edu. (see http://www.mit.edu/~turitsyn/).

}}

 \markboth{IEEE Transactions on Power Systems,~Vol.~, No.~, August ~2014}%
 {Shell \MakeLowercase{\textit{et al.}}: Numerical Polynomial Homotopy Continuation Method to Locate All The Power Flow Solutions}%

 \maketitle

\begin{abstract}



The manuscript addresses the problem of finding all solutions of power flow equations or other similar nonlinear system of algebraic equations. This problem arises naturally in a number of power systems contexts, most importantly in the context of direct methods for transient stability analysis and voltage stability assessment. We introduce a novel form of homotopy continuation method called the numerical polynomial homotopy continuation (NPHC) method that is mathematically guaranteed to find all the solutions without ever encountering a bifurcation. The method is based on embedding the real form of power flow equation in complex space, and tracking the generally unphysical solutions with complex values of real and imaginary parts of the voltage. The solutions converge to physical real form in the end of the homotopy. The so-called $\gamma$-trick mathematically rigorously ensures that all the paths are well-behaved along the paths, so unlike other continuation approaches, no special handling of bifurcations is necessary. The method is \textit{embarrassingly parallelizable} and can be applied to reasonably large sized systems. We demonstrate the technique by analysis of several standard test cases up to the 14-bus system size. Finally, we discuss possible strategies for scaling the method to large size systems, and propose several applications for transient stability analysis and voltage stability assessment.

\end{abstract}


 \maketitle

\graphicspath{{images/}}

\section{Introduction}

Power flow problem is one of the most fundamental and essential problems in power systems. It provides us a ``snapshot" of the power systems in the steady state \cite{Kundur}; hence, the solutions of power flow problem are widely employed in the planning stage. The power flow equations are inherently nonlinear and may have multiple solutions for given set of parameters  \cite{Thorp}. 

In normal operation the system attempts to stay in a single of the multiple solutions, typically the one with the highest levels of voltage. However, characterization of all the solution branches is important for a number of security assessment, planning, and operation processes. For example, the direct methods of transient stability analysis require the knowledge of energy function \cite{Kakimoto:1984kr,Varaiya:1985ig,Chiang:1992ct,Chiang:2011eo}  at the closest unstable equilibrium points (CUEPs) of the system. Similarly, the distance between the normal and the closest unstable equilibrium may be used as the margin to voltage instability \cite{Cutsem}. Recently, the authors of this manuscript have demonstrated the phenomenon of multistability that can arise in distribution grids with power flow reversal \cite{nguyen2014voltage}. The study raised concerns that the transition among equilibria is possible and the system may get entrapped at the undesirable equilibrium in the post-fault recovery.

However, finding all solutions to the power flow problem is, actually, a very challenging task. Several approaches to this problem have been proposed in the literature. These include most importantly the continuation methods \cite{Kundur, Thorp}, and various techniques for finding the closest unstable equilibrium points \cite{Chiang:1992ct,Chiang:2011eo}. The techniques based on the real algebraic geometry methods have been discussed more recently as well \cite{ nguyen2014appearance}. The problem with the purely algebraic methods is their poor scalability with the system size. On the other hand, the continuation methods may fail to find all the solutions in some situations \cite{molzahn2013counterexample}. 

The homotopy continuation methods have been known and applied to the power flow equations for quite some time now, see e.g. \cite{ajjarapu1992continuation}. The basic strategy to solve a system of nonlinear equations using a homotopy continuation method is to start with a new system of nonlinear equations which is easier to solve. Then, one constructs a family of problems through one or more parameters that the system to be solved and the newly constructed systems both are members of. Next, each solution of the new system is tracked towards the original system along the parameters. If the path successfully reaches the original system, then we achieve one solution of the original solution. Performing such a path-tracking for many solutions of the new system towards the original one finally gives many solutions of the original system. Though, in many cases, this approach may be more successful in finding many solutions in comparison with the Newton-Raphson approaches, it does not guarantee to find all the solutions of the original system and can be difficult to apply in situations when the solution tracked by the homotopy experiences a bifurcation. 

When the system of equations is a system of polynomial equations, the homotopy continuation methods performs better than most other approaches. We propose a Numerical Polynomial Homotopy Continuation (NPHC) method in this paper which guarantees to find \textit{all} the solutions. The NPHC method has been only recently applied to various areas in Physics \cite{Mehta:2009,Hughes:2012hg,Mehta:2009zv,Mehta:2011wj,Maniatis:2012ex,Mehta:2012wk,Hauenstein:2012xs,Mehta:2012qr,MartinezPedrera:2012rs,He:2013yk,Mehta:2013fza,Greene:2013ida}, Chemistry, and Mathematical Biology. Due to inevitably expensive computation and limitation of computers, we can solve the power flow problem of $14$-node systems using the NPHC method. Theoretically, solving more than $30$-node system is not completely impossible using the Polyhedral Homotopy which we also introduce in this paper. It should be pointed out that the previous works demonstrating similar methods to the power flow equations either scaled badly with the system size \cite{70552} or were proved not to find all the solutions in practice \cite{ma1993efficient,liu2005toward,molzahn2013counterexample}. Moreover, the NPHC method can store the general form of solutions that only depend on the system parameters, the feature which we will exploit in the power flow problems in a separate article.

The paper is organized as follows. In Section \ref{sec:powerflow}, we first present the power flow problem in rectangular form as a system of polynomial equations. In section \ref{section:pnhc}, we introduce the NPHC method, the Total Degree Homotopy with an explicit example of the $2$-bus system, and Polyhedral Homotopy. In section \ref{sec:simulation}, we give out the details of our computations by first giving a comparison between the Total Degree Homotopy and the Polyhedral Homotopy, demonstrating that the Polyhedral Homotopy significantly reduces the computational efforts compared to the Total Degree Homotopy. Next, we introduce various possible applications of NPHC to power systems in section \ref{sec:apps} toward security analysis/assessment. In appendices \ref{app:4bus} and \ref{app:14bus}, we give details of our computations for the $4$-bus system and IEEE $14$-bus with the NPHC method.

\section{Power flow problem} \label{sec:powerflow}



Traditional approaches for analysis of power flow solutions based on iterative methods like Newton-Raphson, continuation power flow, and their variations \cite{Kundur} are not suitable for identification of all branches of the solution manifold. By construction these methods find only one solution at a time for given set of parameters. There is no systematic way of adjusting the initial conditions that would guarantee that all solutions branches are found. In this work we propose a novel algorithm for identification of all the solutions based on the complex algebraic geometry techniques. This algorithm relies on the polynomial representation of the power flow equations that can be obtained via rewriting the equation in rectangular form. We consider a $n$-bus system which represents either a transmission network or a distribution network. Let bus $1$ be the slack bus. At bus $i$, $ 2\leq i \leq n $, let i) $ P_i $ and $ Q_i $ be active and reactive power consumed or generated at bus $i$ (so $P_i>0$ corresponds to generation); ii) $ V_i=V_{iRe}+jV_{iIm} $ be the rectangular form of bus voltage. The power flow equations can be expressed as follows \cite{nguyen2014appearance}:

\begin{equation} 
\begin{split} \label{eq:pfcom1}
P_i & = \sum\limits_{k=1}^n G_{ik}(V_{iRe}V_{kRe}+V_{iIm}V_{kIm})\\
& +\sum\limits_{k=1}^nB_{ik}(V_{kRe}V_{iIm}-V_{iRe}V_{kIm}); 
\end{split}
\end{equation}

\begin{equation} 
\begin{split} \label{eq:pfcom2}
Q_i &= \sum\limits_{k=1}^n G_{ik}(V_{kRe}V_{iIm}-V_{iRe}V_{kIm})\\
& -\sum\limits_{k=1}^nB_{ik}(V_{iRe}V_{kRe}+V_{iIm}V_{kIm})  
\end{split}
\end{equation}

where $Y_{ik}=G_{ik}+jB_{ik} $ is an entry of the bus admittance matrix, $Y$.


\section{Polynomial Homotopy Continuation Method} \label{section:pnhc}
In this section, we outline the numerical polynomial homotopy continuation (NHPC) method.

The key idea of the NPHC method is to embed the system of real equations \eqref{eq:pfcom1}, \eqref{eq:pfcom2} in a complex space, so that the values of $V_{iRe}$s and $V_{iIm}$s are considered complex in the intermediate steps of the calculations. The main reasoning behind this approach can be explained as follows. The system \eqref{eq:pfcom1},\eqref{eq:pfcom2} does not always have the solutions for real values of  $V_{iRe}$s and $V_{iIm}$s, which correspond to valid values of complex voltage values $V_i$s. Moreover the number of solution may vary as the parameters corresponding to load flows change. However, when $V_{iRe}$s and $V_{iIm}$s are allowed to be complex, the fundamental theorem of algebra guarantees that the solution to \eqref{eq:pfcom1},\eqref{eq:pfcom2} will always exist and their number remains the same. Certainly, complex values of $V_{iRe}$s and $V_{iIm}$s do not make any physical sense and don't correspond to any physical solution. However, as long as the complex values are used only for intermediate calculations, the method produces the valid real solutions in the end.

The NPHC method works as follows \cite{SW:95,Li:2003}: first, one estimates an upper bound on the number of complex solutions of the given system. Then, one writes down the equation or system of equations to be solved in a more general parametric form, based on the upper bound. This process is called \textit{constructing homotopy}. Then, one solves this general system at a point in the parameter space at which its solutions can be easily found. There are various ways of finding such a point of which we will discuss one below. This parameter point is called the starting point, the system at this parameter point is called the start system and the solutions of the start system are called the start solutions. Then, each of the start solutions are tracked in the parameter space from the starting point to the point in the parameter space corresponding to the original system. 

More specifically, let us consider a system of multivariate polynomial equations $P(x)=0$, where $P(x)=(p_{1}(x),\dots,p_{m}(x))$ and $x=(x_{1},\dots,x_{m})$, which is known to have isolated solutions. 

As the first step for the NPHC method, we need to estimate an upper bound on the number of solutions of this system. A popular choice for an upper bound comes from the well-known \textit{Classical B\'ezout Theorem}, which asserts that for a system of $m$ polynomial equations in $m$ variables the maximum number of solutions in $\mathbb{C}^{m}$ is $\prod_{i=1}^{m}d_{i}$, where $d_{i}$ is the degree of the $i$th polynomial. The Classical B\'ezout Bound (CBB), is exact if the coefficients take generic values.

Then we can construct a homotopy as 
\begin{equation}
H(x,t)=\gamma(1-t)Q(x)+t\; P(x)=0,
\end{equation}
where $Q(x)$ is another system of polynomial equations, $Q(x)=(q_{1}(x),\dots,q_{m}(x))$, i.e., the start system,
with the following properties: 
\begin{enumerate}
\item All the complex (which also include all the real) solutions of $Q(x)=H(x,0)=0$ 
are known or can be easily obtained.
\item The number of solutions of $Q(x)=H(x,0)=0$ is equal to an estimated
number (or an upper bound) of the solutions for $P(x)=0$, i.e., in our case, equal to the CBB of $P(x)=0$.
\item The solution set of $H(x,t)=0$ for $0\le t\le1$ consists of a finite
number of smooth paths, called homotopy paths, each parametrized by
$t\in[0,1)$. 
\item Every isolated solution of $H(x,1)=P(x)=0$ can be reached by some
path originating at a solution of $H(x,0)=Q(x)=0$. 
\end{enumerate}
We can then track each path corresponding to each solution of
$Q(x)=0$ from $t=0$ to $t=1$ and reach $P(x)=0=H(x,1)$. By implementing
an efficient path tracker algorithm, e.g., the corrector-predictor method, we can get all the isolated solutions
of $P(x) = 0$. 

\noindent\textbf{The $\gamma$-trick:}
Here, $\gamma=e^{i\theta}$ with $\theta\in\mathbb{R}$ is a generic random number. It is shown \cite{Morgan1987101,SW:95,Li:2003} that for all but a finite number of angles $\theta$, $H^{-1}(x,0)$ consists of smooth paths over $[0, 1)$ and every isolated solution of $P(x)= 0$ has a path converging to it. Furthermore, if $m^*$ is the multiplicity of an isolated solution $x^*$, then $x^*$ has exactly $m^*$ paths converging to it. And all the paths are strictly increasing in $t$.

In other words, for a generic value of $\gamma$ each path is well-behaved for $t\in[0,1)$. This trick, called the $\gamma$-trick, in fact makes sure that there is no singularity or bifurcation along the paths and that the NPHC method then guarantees to find all the solutions with any start system which satisfies the above four criteria.

Note that the numerical homotopy continuation approach can be applied even to non-algebraic equations. However, there exist several difficulties. In particular, in the non-algebraic systems case, this method may not guarantee to find all the solutions. Hence, the method may not always be a primary candidate method to solve a set of non-algebraic equations. However, due to the reasons such as the applicability of the $\gamma$-trick to polynomial systems, the algebraic geometry ways of finding upper bounds on the number of solutions, etc., this method works exceptionally well for polynomial systems.

The homotopy constructed using the CBB is called the \textit{Total Degree Homotopy} and the start system $Q(x)=0$ can be taken for example as 
\begin{equation}
Q(x)=\left(\begin{array}{c}
x_{1}^{d_{1}}-1\\
x_{2}^{d_{2}}-1\\
\dots\\
x_{m}^{d_{m}}-1
\end{array}\right)=0,\label{eq:Total_Degree_Homotopy}
\end{equation}
where $d_{i}$ is the degree of the $i^{th}$ polynomial of the original system $P(x)=0$. Solving Eq.~(\ref{eq:Total_Degree_Homotopy}) is obviously easy. Moreover, the total number of start solutions is $\prod_{i=1}^{m}d_{i}$, all of which are non-singular.

\subsection{Example: The $2$-bus system}
We now take a simple system to explain the steps of the NPHC method. The system we consider is the one load - one infinite bus system with purely inductive line. We choose this setting to illustrate how the method works on the systems with different equations order. The parameters used for the simulation ($V_{1}= 1+j 0$, $P_{2}=0.1$, $Q_{2} = 0.1$ and $Y=- j 6.67$) result in the following algebraic system:
\begin{equation}
f(V) = \left(\begin{array}{c}
0.1+6.667 V_{2Im}\\
0.1+6.667 V_{2Im}^2-6.667 V_{2Re}+6.667 V_{2Re}^2
\end{array}\right)=0,
\end{equation}
where $V_{2Im}$ and $V_{2Re}$ are the variables. Since the first equation is a univariate linear equation, it can be easily solved for $V_{2Im}$. Then, substituting the solution of $V_{2Im}$ in the second equations, we can also find both the solutions of the second equation in $V_{2Re}$. However, we solve it with the NPHC method pretending that the solutions of this system are not known.

In the first step, we estimate an upper bound on the number of complex solutions using the CBB. Since the degrees of the first and the second polynomials are $1$ and $2$, respectively, the CBB is $1\times 2 = 2$.  Then, we create a new system, say $g(x)$, as below:
\begin{equation}
g(V) = \left(\begin{array}{c}
V_{2Im}-1\\
V_{2Re}^2-1
\end{array}\right)=0.
\end{equation}
It is easy to see that $g(V)$ has the same number of roots as the CBB, i.e., $2$ solutions, $(1,1)$ and $(1,-1)$. We now construct a homotopy between $f(V)$ and $g(V)$ as
\begin{equation}
H(V,t)=\gamma(1-t)g(V)+t\; f(V)=0,
\end{equation}
where we pick a generic complex number $\gamma$. The homotopy paths can now be tracked from each of the solutions $(1,1)$ and $(1,-1)$ of $g(V)=0$ from $t=0$ to $t=1$ using the predictor-corrector method. In the end, for this system, both the solution-paths converge to $t=1$, and we get the following solutions: $(-0.015, 0.016)$ and $(-0.015, 0.985)$. Note that these solutions can be refined further to the desired precision.

\begin{figure}[ht]
    \centering
    \includegraphics[width=0.9 \columnwidth]{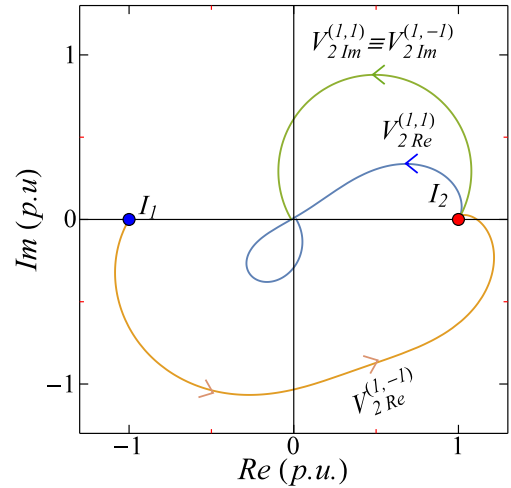}
	\caption{Homotopy paths of the $2$-bus system}
    \label{fig:2bustrajectory}
\end{figure}

Figure \ref{fig:2bustrajectory} visualizes how the power flow solutions, ($V_{2Im}$ and $V_{2Re}$), evolve as $t$ increases from $0$ to $1$ as indicated by the arrows directions in complex plane. Each trajectory is plotted in different color, that corresponds to the homotopy path of either $V_{2Im}$ or $V_{2Re}$. The superscripts of $(1,1)$ or $(1,-1)$ denote the start solutions, i.e. the two solutions of the start system or $g(V)=0$, which associates with either the large blue dot, $I_1$, or large red dot, $I_2$, on the real axis. One can see that each trajectory starts from either $I_1$ or $I_2$ lying on the real axis at $t=0$; then from there, it moves away to complex values of $V_{2Im}$ and $V_{2Re}$ which are, obviously, unphysical; finally, it finishes at the real axis where $t=1$ indicating the power flow solution which is a real number. Therefore, intermediate values of $V_{2Im}$ and $V_{2Re}$ are complex numbers, this illustrates the key idea of NPHC. One can also can observe that, for the $2$-bus system, commencing with two start solutions of $g(V)=0$, $(1,1)$ and $(1,-1)$, the homotopy paths converge to two solutions to the power flow problem, i.e $(-0.015, 0.016)$ and $(-0.015, 0.985)$, respectively. 

\subsection{The Polyhedral Homotopy}

Though the total degree homotopy is the simplest one in that the CBB is easy to compute and the corresponding start system is also easy to come up with and solve, polynomial equations arising
from real-life problems are \textit{sparse}, i.e., there are usually only a few terms in each polynomial equation of the system. In such spase systems, the actual number of solutions is much smaller than the CBB. Hence, a lot of computational effort is wasted in tracking paths that eventually turn out to be non-solutions. In order to resolve this issue, we also use a tighter upper bound on the number solutions, called the Bernstein-Khovanskii-Kushnirenko (BKK) bound. Because of the more technical nature of the computation of the BKK bound, the details are explained in Appendix \ref{app:BKK}. 

The homotopy in which the start system is  on the BKK bound is called \textit{Polyhedral Homotopy}. The advantage of this homotopy compared to the Total Degree Homotopy is that one has to track less number of paths. The disadvantage of the Polyhedral Homotopy is that solving the start system based on the BKK bound may not be as easy as in the case of the Total Degree Homotopy. However, in many problems including ours, the time to solve the start system for the Polyhedral Homotopy may be negligible compared to the time it saves to track the extra paths compared the the Total Degree Homotopy.

There are several sophisticated numerical packages such as \textsf{Bertini} \cite{BHSW06}, PHCpack~\cite{Ver:99},
PHoM~\cite{GKKTFM:04}, HOMPACK~\cite{MSW:89} and HOM4PS2 \cite{GLW:05,Li:03} well-equipped to solve multivariate polynomial systems. In our work, we have used HOM4PS2 for the polyhedral homotopy computation and Bertini for the total degree homotopy computation. 

\subsection{Salient Features of the NPHC Method}
In this subsection,we list out salient features of the NPHC method:
\begin{enumerate}
\item The NPHC method, unlike the Newton-Raphson and many other local numerical methods, guarantees to find all the isolated complex solutions of a given algebraic system of equations. Since the complex solutions also include real solutions, which are most of the times the physically relevant ones, in the end we obtain all the real solutions. The non-real solutions at least in the power-flow equations have no physical meaning. This means that a lot of computational effort is wasted in finding these solutions. On the other hand, at the moment, a method which finds only \textit{all} the real solutions is not available. Such a method, however, will be one of the most significant advancements not only in the power-flow areas but also in Mathematics itself.

\item The NPHC method is known as being \textit{embarrassingly parallelizable} because each solution-path can be tracked independent of all others. Moreover, as all the paths are described by the same differential equation, it can be very efficiently implemented on modern Single Instruction Multiple Data (SIMD) architecture like Graphical Processing Units that currently offer significantly more computing power for a given cost of hardware. 
\item Unlike the symbolic methods, the NPHC being a purely numerical method does not need to have systems with only rational coefficients.
\item Unlike the methods based on the Newton-Raphson method, the PNHC method finds all the isolated singular solutions same as non-singular solutions.
\item One can extend the application of the NPHC method to systems with parameters by tracking solution-paths over the parameters. Thus, after solving the parametric system at one generic parametric point, one can then easily obtain solutions at any other parameter-point of the system. Solving the system at many parameter points can then help to visualize how the system varies with the parameters. We will exploit this feature of the NPHC method, called the parameter homotopy, in another paper.
\end{enumerate}

\section{Results} \label{sec:simulation}
In Table \ref{table:comparison_CBB_BKK}, we record the CBB and BKK bounds for the $10$, $11$, $13$, $14$-bus systems \cite{taylor1994power,DurgaPrasad1990,IEEETranstestcase}, and the $12$-bus one is built based on IEEE 14-bus test case, as well as an example system for the $5$-bus case given in \cite{molzahn2013counterexample}. The systems of $11$ and $13$ buses are ill-conditioned power systems.

Since all the $2(n-1)$ power flow equations are quadratic, the CBB is $2^{2(n-1)}$. The BKK bound however is significantly smaller than the CBB for all $n$ except for $n=5$ for which the CBB and BKK bound are equal. Due to the limited computational resources available to us, we can go up to $14$ bus system so far.

Using the Polyhedral Homotopy, the systems for $n\leq 10$ were solved in less than one hour and $10<n\leq 13$ were solved in around $6,7$ and $9$ hours on a desktop. For $N=14$, we employed a computing cluster with 64 processors that approximately took around $2$ hours.

For all the cases, each equation was zero with $10^{-10}$ tolerance at each solution. Each real solution was identified as a solution with zero imaginary part with tolerance $10^{-10}$. Both the Total Degree Homotopy and the Polyhedral Homotopy give the same complex as well as real solutions in the end as they should, except that the latter is far more efficient specially as the systems grow big. Power flow solutions of the $4$-bus and IEEE $14$-bus systems are listed in Appendix.

\begin{table}[ht]
    \caption{Comparison between the CBB and the BKK bounds for the IEEE test systems. The BKK bound is significantly smaller than the CBB in all the cases, except $n=5$ which is taken from ~\cite{molzahn2013counterexample}.}
    \label{table:comparison_CBB_BKK}
    \centering
   \begin{tabular}{|c |c |c|}
    \hline
    \textbf{$n$-bus case} & CBB & BKK \\
               \hline
    $5$ & $256$ & $256$ \\           
    $10$ & $262,144$ & $144,384$ \\
    $11$ & $1,048,576$ & $770,048$ \\
    $12$ & $4,194,304$ & $3,080,192$ \\
    $13$ & $16,777,216$ & $12,320,768$ \\
    $14$ & $67,108,864$ & $49,283,072$ \\
    \hline
    \end{tabular}
\end{table}

\subsection{The $5$-bus Problem}
Among the cases we have worked on, the $5$-bus case needs a special attention as this is the problem used as a counter-example of the claim a previous work made: in \cite{ma1993efficient} a continuation based method was presented claiming to find all the solutions of the power-flow problem. However, in \cite{molzahn2013counterexample}, the claim was disproved by showing the method not able to find all the solutions in the above mentioned $5$-bus case.

With the NPHC method, using both the CBB and BKK bound, we have obtained all the $10$ real solutions of this system demonstrating the reliability of our method.

\section{Possible applications in power systems} \label{sec:apps}
NPHC is not only useful for power flow problem but also can be applied to a number analysis/assessment applications that are discussed below.

\subsection{Transient stability assessment}
In the context of power systems the NPHC approach can be most naturally applied for identification of the Closest Unstable Equilibrium Points (CUEP) for energy methods. Energy methods is probably the most powerful technique for rigorous dynamic security assessment of power systems. Based on classical Lyapunov function approach they allow to certify the nonlinear stability of the system without running the extensive numerical simulations of the large-scale dynamics of the power system. This certification is accomplished by comparison of the effective ``energy'' of the initial state of the system to the energy of an unstable equilibrium point close to the equilibrium state of the system. Both the stable and unstable equilibrium points are extrema of the energy functional with stable equilibrium corresponding to its minimum, while the CUEPs corresponding to the saddle nodes. Most of the research efforts in the energy methods field have targeted the problem of identification of the CUEPs that is an essential and the most time consuming step of the method \cite{Chiang:1989gn,Liu:1997co,Lee:2003hg,Lee:2004ch,Lee:2003is,Chen:2009gs,Yorino:2013do,Hristov:2013kc}.

The energy functions have been developed for large number of power system models including the very sophisticated ones. In the simplest situation of swing equations defined for purely inductive networks, the extrema of the energy functions are defined by the power-flow like equation of the form
\begin{equation}
 P_i = \sum_j B_{ij} |V_i| |V_j| \sin(\delta_i - \delta_j),
\end{equation}
where $B_{ij}$ is the Kron-reduced susceptance matrix of the power grid, $V_i$ are the regulated voltage levels on the generation buses, and $\delta_i$ are the rotor angles of the synchronous generators. This system can be transformed into polynomial form via a change of variables $x_i = \sin\delta_i$ and $y_i = \cos\delta_i$ and supplementary equations $x_i^2 + y_i^2 = 1$. The NHPC method guarantees to find all the unstable equilibrium points with zero missing rate. Although many of the equilibrium points found by the NHPC method won't be the interesting Type-1 ones, it is to our knowledge the only method that is proven to find all the equilibria and thus provides mathematical guarantees for stability. 

\subsection{Voltage stability}
For voltage stability analysis, a common security indicator, characterizing the proximity to instability is the distance between the stable equilibrium, $S$, which is normally the operating point, and the closet unstable one, $U$ \cite{Cutsem}. For traditional systems it can be visualized on a typical $PV$ curve. In most of the practically relevant situations, the two equilibria $S$ and $U$ correspond to the same level of active power $P$, but different levels of voltage. The $S$ point normally lies on the upper branch, whereas $U$ is on the lower one. At the nose tip point corresponding to the maximal loadability, the system exhibits saddle-node bifurcation (SNB), and the two equilibria converge and disappear resulting in the voltage collapse phenomena in real life. In practical applications, identification of the closest unstable equilibrium may be a difficult problem especially in multi-dimensional setting where multiple directions in the power space are relevant. This situation is likely to be more and more relevant in the future as the penetration of the unpredictable renewable resources increases. Moreover, the number of solutions to power flow equations may be very large, and traditional continuation type of techniques may not guarantee to find the closest of the solutions \cite{Kundur}. The NPHC technique provides a list of all possible solutions that can be then ranked by some distance measure, preferably reflecting the probability of the system moving in a given direction of the load level space. The corresponding sorted list of closest equilibria will not only provide the estimate for proximity, but will also contain information about the most dangerous directions in the phase space. This information could be naturally used for choosing the optimal response actions, such as generator redispatch based control sensitivities \cite{OverbyeDmarco1991,Overbye1991}.


\subsection{Multistability phenomenon}
For large levels of penetration of distributed generation the distribution power grids may enter the power flow reversal regime. It has been shown recently that the number of solution of power flow equations may increase quite dramatically. Recent studies have indicated that many of those solutions are actually stable \cite{nguyen2014appearance,nguyen2014voltage} so that the system may get entrapped in them if no adequate undervoltage protection is installed on low voltage parts of the grid. One of the most adverse consequences of the multistability phenomenon is that the undesirable low voltage stable equilibrium may prevent the normal recovery of the system after transient disturbances and faults. This phenomenon is similar to Fault-Induced Delay Voltage Recovery (FIDVR), but instead of just delaying the recovery, the entrapment completely prevents it. Instead of returning to the normal state, the system may remain in a low-voltage emergency state. Inappropriate counter-measurements then may lead to either partial or complete blackouts. To avoid such situations, the security assessment procedures need to incorporate information about the dynamic characteristics of the new solutions that necessitates finding of all possible solutions to the power flow problem. 

\subsection{Applications to large scale systems}

As aforementioned, the main drawback of NPHC and most of the other similar techniques is its poor scalability with the system size. The technique can be applied only to small size systems of no more than $15$ or so buses. The standard way of avoiding this limitation is to use the small equivalents of the real large-scale systems. The appropriate strategy for voltage stability type problems would be to combine the partitioning of a system in a set of weakly interacting sub-systems, and consequent reduction of the outer part of a given sub-system to some simple polynomial response to voltage drop. In this case the voltage stability problem could be studied via small size models and all the local branches of the solution could be identified with limited computational resources.

This strategy will likely work well for voltage stability studies. Unlike frequency, the voltage is generally a local quantity, that is influenced by local reactive power injections. The reason is that in normal situations, the voltage difference between two adjacent buses is insignificant, so reactive power cannot be transmitted over long distances. Hence, the voltage control effect of individual voltage control source is also regional. For the static voltage control problem, it is reasonable to reorganize and divide the network into sub-networks by either applying the concept of sphere of influence (SOI) proposed in \cite{JEET} or echelon approach as in \cite{monvolilic}. The size of each sub-network must be small enough to be solved with NPHC. In this sense, NPHC could be applied to explore the local solution manifolds.

\subsection{Static solution boundary}
The approach could be also potentially applied for identification of solvability and stability boundaries of the system used for planning and contingency analysis. This boundaries are used for example in preventive control which restores the solvability of the system; the static solution boundary is then used as solvable boundary. For an appropriate size system, NPHC could be used to characterize the solution manifold by solving the following set of polynomial equations:

\begin{equation} 
f(x,\lambda) = 0 \nonumber
\end{equation}
\begin{equation} \label{eq:hk}
f_x(x,\lambda)v = 0
\end{equation}
\begin{equation} 
v^tv = 1\nonumber
\end{equation}

Here $v$ is the right eigenvector of the Jacobian matrix corresponding to its zero eigenvalue. If the total number of parameters is equal to $k$, so that $\lambda \in \mathbb{R}^k$, and both the vectors $x$ and $v$ have $2(n-1)$ components, so that $x,v\in \mathbb{R}^{2n-2}$, the total number of variables in this equation is $k+4n-4$. The total number of equations on the other hand is $4n-3$, so the system (\ref{eq:hk}) describes the $k-1$ dimensional manifold in the parameter space $\lambda$.









\section{Conclusion}
In this work, we introduced a novel method, NPHC, to find all the solutions to power flow problem (or other polynomial systems of equations arising in power system). The main advantage of this method is its mathematically certified ability to find all the solutions to polynomial equations. Unlike other homotopy techniques, the homotopy paths never encounter bifurcations and admit much easier implementation and parallelization. The main downside of the method is its poor scalability to large scale systems, which is common to all the homotopy techniques aiming to find all the solutions to the power flow problem.

Various test cases consisting up to 14 buses have been solved to demonstrate NPHC performance. System partitioning may be used to circumvent the limitation of system size and enable application of NPHC to realistic power system models. We suggested/proposed several security analysis/assessment applications of NPHC to power system, i.e. transient stability and voltage stability assessments, analysis multistability phenomenon, voltage monitoring in large scale systems, and charcterization of solvability and stability boundaries.

In the future, we plan to combine the NPHC with system partitioning and reduction techniques and explore its scalability to large scale systems. Moreover, extending NPHC to produce the parameter dependencies of the solutions may lead to new generation of tools useful for planning, real-time security assessment, and nonlinear control approaches. 

On mathematical side, we are also developing one more and rather powerful extension of the NPHC called the numerical discriminant variety method. This method will efficiently provide the boundaries inside the parameter space at which the number of real solutions changes. These boundaries are essential for power system operators in finding optimal corrective and preventive controls without linearizing and approximating the system, and also helps in planning stages.

\section{Acknowledgement}
Dhagash Mehta was supported by a DARPA YFA award. Hung Nguyen and Konstantin Turitsyn thank the MIT/Skoltech and Masdar Initiative, as well as the Vietnam Education Foundation for their support. 

\appendices

\section{The $4$-bus system}\label{app:4bus}

\begin{figure}[ht]
    \centering
    \includegraphics[width=0.65 \columnwidth]{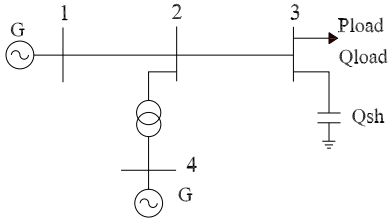}
	\caption{The $4$-bus network}
    \label{4bus}
\end{figure}
The $4$-bus system configuration is shown in Figure. \ref{4bus} \cite{Cutsem}.
\begin{table}[ht]
    \caption{The $4$-bus system branch data}
    \centering
    \label{branchdata4bus}
    \begin{tabular}{|c |c |c|}
    \hline
    \textbf{From bus} & \textbf{To bus} & \textbf{Line impedance $(p.u.)$} \\
               \hline
    $1$ & $2$ & $j\,0.20$ \\
    $2$ & $3$ & $j\,0.05$  \\
    $2$ & $4$ & $j\,0.03$ \\
    \hline
    \end{tabular}
\end{table}

\begin{table}[ht]
    \caption{The $4$-bus system bus data}
    \centering
    \label{busdata4bus}
    \begin{tabular}{|c |c |c|c |c |c|c|}
    \hline
    \textbf{Bus} & \textbf{Bus type}& \textbf{Vsch} & \textbf{Pload}& \textbf{Qload }& \textbf{Pgen }& \textbf{Qsh}\\
     & & \textbf{$(p.u.)$} & \textbf{ $(p.u.)$}& \textbf{ $(p.u.)$}& \textbf{ $(p.u.)$}& \textbf{ $(p.u.)$}\\
               \hline
    $1$ & $3$ & $1.08$ &$NA$ & $NA$ & $NA$& $0$ \\
    $2$ & $1$ & $NA$ &$0$ & $0$ & $0$ & $0$\\
    $3$ & $1$ & $NA$ &$0.3$ & $0.1$ & $0$& $0.5$ \\
    $4$ & $2$ & $1.05$ &$0$ & $0$ & $0.1$ & $0$\\
    \hline
    \end{tabular}
\end{table}

In Table \ref{busdata4bus}, bus types of $1$, $2$, $3$ denote $PQ$ bus, $PV$ bus, and slack bus, respectively. $Vsch$ is the voltage scheduled at PV buses and slack bus. $Pload$ and $Qload$ represent the consumption levels at corresponding $PQ$ buses. $Pgen$ is the generation level specified at $PV$ bus. $Qsh$ is the fixed amount of reactive powers that fix shunts produce.

For the $4$-bus system, NPHC returns $4$ distinct solutions as shown in Table \ref{solution4bus}.

\begin{table}[ht]
    \caption{Power flow solutions of the $4$-bus test case}
    \centering
    \label{solution4bus}
    \begin{tabular}{|c |c | c | c |c|}
    \hline
     \textbf{Bus}& \textbf{$1$st solution} & \textbf{$2$nd solution}& \textbf{$3$rd solution} &\textbf{$4$th solution}\\
               \hline
    \textbf{$2$} & $-0.5-j0.4 $ & $-0.61-j0.4$  & $0.61-j0.4$& $0.92-j0.4$\\
    \textbf{$3$} & $-0.39-j0.01$ & $ -0.61-j0.15 $  &$ 0.02-j0.26$& $ 0.82-j0.52$\\
    \textbf{$4$} & $-0.79-j0.69$ & $-0.86-j0.61$  & $0.9-j0.54$& $0.98-j0.39$\\
    \hline
    \end{tabular}
\end{table}

\section{IEEE $14$-bus test case}\label{app:14bus}

\begin{figure}[ht]
    \centering
    \includegraphics[width=0.6 \columnwidth]{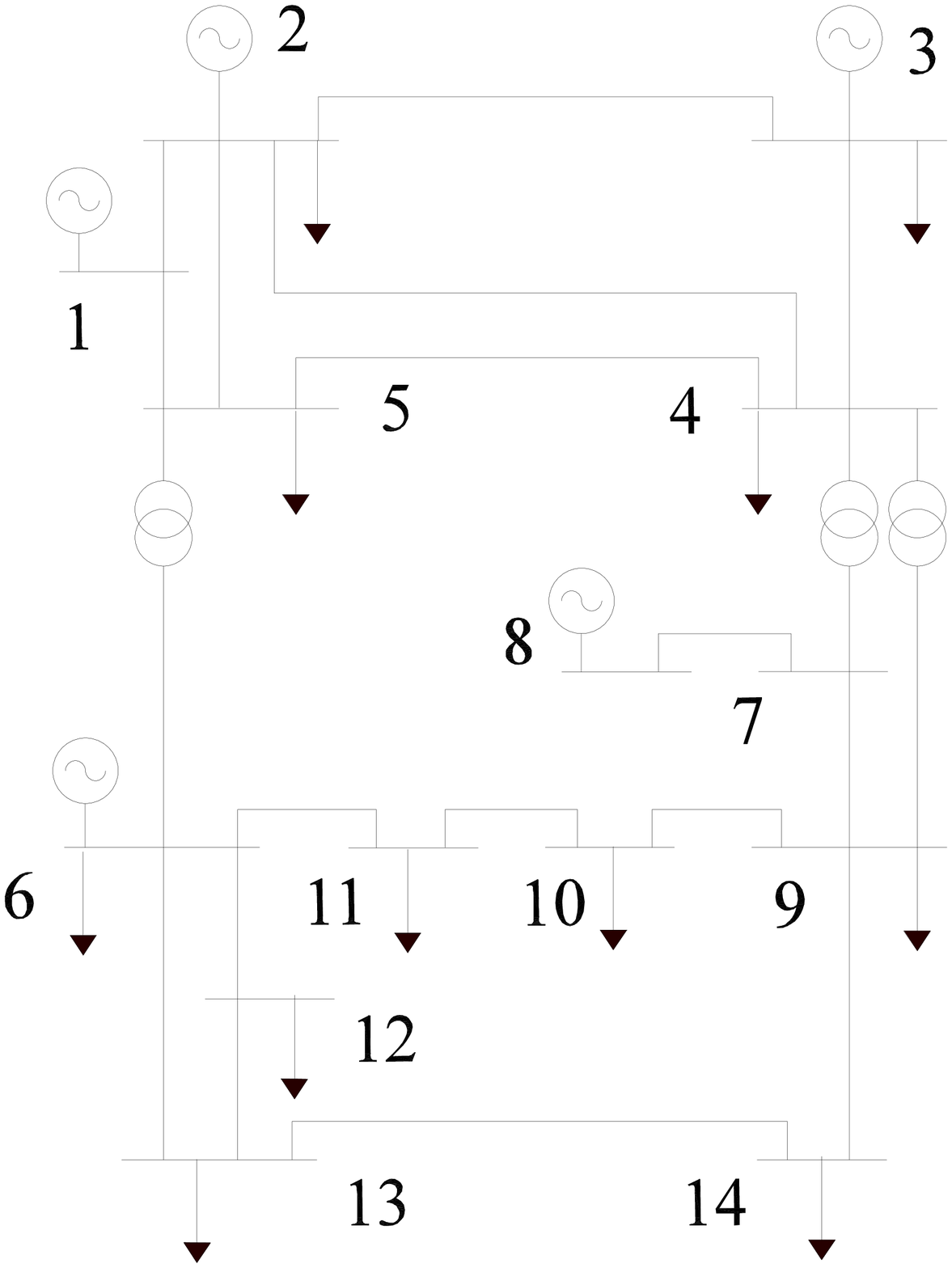}
	\caption{IEEE $14$-bus test case}
    \label{14bus}
\end{figure}

IEEE $14$-bus test case is depicted in Figure. \ref{14bus} \cite{IEEETranstestcase} and the branch data as well as bus data are listed in Table \ref{branchdata14bus} and \ref{busdata14bus}, respectively.

\begin{table}[ht]
    \caption{IEEE $14$-bus test case branch data}
    \centering
    \label{branchdata14bus}
   \begin{tabular}{|c |c |c|c|}
    \hline
    \textbf{From bus} & \textbf{To bus} & \textbf{Line impedance $(p.u.)$} & \textbf{Shunt $(p.u.)$}\\
               \hline
    $1$ & $2$ & $0.01938+j\,0.05917$ & $0.0528$\\
    $1$ & $5$ & $0.05403+j\,0.22304$  &$0.0492$\\
    $2$ & $3$ & $0.04699+j\,0.19797$ &$0.0438$\\
    $2$ & $4$ & $0.05811+j\,0.17632$ & $0.034$\\
    $2$ & $5$ & $0.05695+j\,0.17388$  &$0.0346$\\
    $3$ & $4$ & $0.06701+j\,0.17103$ &$0.0128$\\
    $4$ & $5$ & $0.01335+j\,0.04211$ &$0$\\
    $4$ & $7$ & $j\,0.20912$ &$0$\\
    $4$ & $9$ & $j\,0.55618$ &$0$\\
    $5$ & $6$ & $j\,0.25202$ &$0$\\
    $6$ & $11$ & $0.09498+j\,0.1989$ &$0$\\
    $6$ & $12$ & $0.12291+j\,0.25581$ &$0$\\
    $6$ & $13$ & $0.06615+j\,0.130279$ &$0$\\
    $7$ & $8$ & $j\,0.17615$ &$0$\\
    $7$ & $9$ & $j\,0.11001$ &$0$\\
    $9$ & $10$ & $0.03181+j\,0.0845$ &$0$\\
    $9$ & $14$ & $0.12711+j\,0.27038$ &$0$\\
    $10$ & $11$ & $0.08205+j\,0.19207$ &$0$\\
    $12$ & $13$ & $0.22092+j\,0.19988$ &$0$\\
    $13$ & $14$ & $0.17093+j\,0.34802$ &$0$\\
    \hline
    \end{tabular}
\end{table}

\begin{table}[ht]
    \caption{IEEE $14$-bus test case bus data}
    \centering
    \label{busdata14bus}
    \begin{tabular}{|c |c |c|c |c |c|c|}
    \hline
    \textbf{Bus} & \textbf{Bus type}& \textbf{Vsch} & \textbf{Pload}& \textbf{Qload }& \textbf{Pgen }& \textbf{Qsh}\\
     & & \textbf{$(p.u.)$} & \textbf{ $(p.u.)$}& \textbf{ $(p.u.)$}& \textbf{ $(p.u.)$}& \textbf{ $(p.u.)$}\\
               \hline
    $1$ & $3$ & $1.06$ &$0$ & $0$ & $2.324$& $0$ \\
    $2$ & $2$ & $1.045$ &$0.217$ & $0.127$ & $0.4$ & $0$\\
    $3$ & $2$ & $1.01$ &$0.942$ & $0.19$ & $0$ & $0$ \\
    $4$ & $1$ & $NA$ &$0.478$ & $-0.039$ & $0$ & $0$\\
    $5$ & $1$ & $NA$ &$0.076$ & $0.016$ & $0$& $0$ \\
    $6$ & $2$ & $1.07$ &$0.112$ & $0.075$ & $0$ & $0$\\
    $7$ & $1$ & $NA$ &$0$ & $0$ & $0$& $0$ \\
    $8$ & $2$ & $1.09$ &$0$ & $0$ & $0$ & $0$\\
    $9$ & $1$ & $NA$ &$0.295$ & $0.166$ & $0$& $0.19$ \\
    $10$ & $1$ & $NA$ &$0.09$ & $0.058$ & $0$ & $0$\\
    $11$ & $1$ & $NA$ &$0.035$ & $0.018$ & $0$& $0$ \\
    $12$ & $1$ & $NA$ &$0.061$ & $0.016$ & $0$ & $0$\\
    $13$ & $1$ & $NA$ &$0.135$ & $0.058$ & $0$& $0$ \\
    $14$ & $1$ & $NA$ &$0.149$ & $0.05$ & $0$ & $0$\\
    \hline
    \end{tabular}
\end{table}

IEEE $14$-bus test case has $30$ distinct solutions. Here we present only the first $2$ solutions in Table \ref{solution14bus}.

\begin{table}[ht]
    \caption{Power flow solutions of IEEE $14$-bus test case}
    \centering
    \label{solution14bus}
    \begin{tabular}{|c|c|c|}
    \hline
    \textbf{Bus} & \textbf{$1$st solution (p.u.)} & \textbf{$2$nd solution (p.u.)}\\
               \hline
    \textbf{$2$} & $ 0.17-1.03j  =1.04\angle{-1.41}$ & $  0.53-0.9j  =1.04\angle{-1.04}$  \\
    \textbf{$3$} & $ -0.83+0.58j   =1.01\angle{2.53}$ & $  -1+0.17j  =1.01\angle{2.97}$  \\
    \textbf{$4$} & $  -0.26-0.06j   =0.27\angle{-2.91}$ & $  -0.2-0.11j  =0.23\angle{-2.64}$  \\
    \textbf{$5$} & $  -0.11-0.13j   =0.17\angle{-2.27}$ & $  -0.03-0.14j  =0.14\angle{-1.78}$  \\
    \textbf{$6$} & $  -1.03+0.29j  =1.07\angle{2.87}$ & $  -1.04+0.26j  =1.07\angle{2.90}$  \\
    \textbf{$7$} & $  -0.71+0.07j   =0.71\angle{3.04}$ & $  -0.69  =0.69\angle{3.14}$  \\
    \textbf{$8$} & $  -1.08+0.11j   =1.09\angle{3.04}$ & $  -1.09  =1.09\angle{3.14}$  \\
    \textbf{$9$} & $   -0.71+0.11j    =0.72\angle{2.99}$ & $  -0.7+0.05j  =0.70\angle{3.07}$  \\
    \textbf{$10$} & $   -0.76+0.15j    =0.77\angle{2.95}$ & $  -0.75+0.09j  =0.76\angle{3.02}$  \\
    \textbf{$11$} & $  -0.89+0.22j   =0.92\angle{2.90}$ & $  -0.89+0.17j =0.91\angle{2.95}$  \\
    \textbf{$12$} & $  -0.99+0.3j   =1.03\angle{2.85}$ & $  -0.99+0.26j =1.02\angle{2.88}$  \\
    \textbf{$13$} & $  -0.96+0.28j   =1\angle{2.86}$ & $  -0.97+0.24j  =1.00\angle{2.90}$  \\
    \textbf{$14$} & $  -0.79+0.2j   =0.81\angle{2.89}$ & $ -0.79+0.15j  =0.80\angle{2.95}$  \\
    \hline
    \end{tabular}
\end{table}

\section{The BKK Bound}\label{app:BKK}
In this Appendix, we introduce the BKK bound of the solutions of a system of multivariate polynomial equations. We first need to state the definitions of 
Newton Polytopes and Mixed Volume. 

At each step, we take an example to make the concept clearer. Consider a two-variable system
\begin{eqnarray}
f_{1}(x,y) & = & 1+ax+bx^{2}y^{2}=0,\nonumber \\
f_{2}(x,y) & = & 1+cx+dy+exy^{2}=0,\label{example_system_for_mixed_vol}.
\end{eqnarray}
Here, $x$ and $y$ are complex variables and $a,b,c,d,e$ are complex coefficients. Since the degrees of these polynomials are $4$ and $3$, the CBB of this system is $4\cdot 3=12$.

If for every pair of points within a set $X$, every point on the straight line segment that joins them is also within $X$, then $X$ is called a convex set and the minimal convex set containing $X$ is called the convex hull of $X$. 

The support $S$ of a polynomial is the vector consist of the exponents of each monomial. e.g., the supports of the above two equations are $S_{1}=\{(0,0),(1,0),(2,2)\}$ and $S_{2}=\{(0,0),(1,0),(0,1),(1,2)\}$, respectively.

The convex hull of support $S$ of a polynomial, say $Q=\mbox{conv}(S)$, is called the Newton polytope of the polynomial. e.g., the Newton
polytope for $f_{1}(x,y)$ is $Q_{1}=\mbox{conv}(S_{1})=\{(0,0),(1,0),(1,1),(2,2)\}$ and for $f_{2}(x,y)$ it is $Q_{2}=\mbox{conv}(S_{2})=\{(0,0),(1,0),(0,1),(1,1),(1,2)\}$.

A Minkowski sum of any two convex sets is defined as
\begin{equation}
Q_{1}+Q_{2}=\{q_{1}+q_{2}\colon q_{1}\in Q_{1},q_{2}\in Q_{2}\}.
\end{equation}
The Minkowski sum of two Newton polytopes then corresponds to multiplying the corresponding polynomials algebraically.

Now, since, the $m$-dimensional volume of a simplex having vertices $v_{0},v_{1},\dots,v_{m}$, is
\begin{equation}
\mbox{Vol}_{m}(\mbox{conv}(v_{0},\dots,v_{m}))=\frac{1}{m!}|\det[v_{1}-v_{0},\dots,v_{m}-v_{0}]|,
\end{equation}
one can show that the $m$-dimensional volume $\mbox{Vol}_{m}(\lambda_{1}Q_{1}+\dots+\lambda_{m}Q_{m})$, where $0\ge\lambda_{i}\in\mathbb{R}$, is a homogeneous polynomial of degree $m$ in variables $\lambda_{i}$. The mixed volume of convex
polytopes $Q_{1},\dots,Q_{m}$, denoted $M(Q_{1},\dots,Q_{m})$, is defined as the coefficient of $\lambda_1\cdots\lambda_m$
in $\mbox{Vol}_{m}(\lambda_{1}Q_{1}+\dots+\lambda_{m}Q_{m})$, then
\begin{equation}
M(Q_{1},\dots,Q_{m})=\sum_{i=1}^{m}(-1)^{m-i}\mbox{ Vol}_{m}(\sum_{j\in\Omega_{i}^{m}}Q_{j}),
\end{equation}
where the inner sum is a Minkowski sum of polytopes and $\Omega_{i}^{m}$ are the combinations of $m$-objects (i.e., $m$-dimensional geometrical objects made of $m$-simplices) taken $i$ at a time. Obviously, the mixed volume is always an integer for a system
of polynomials. e.g., in our running example (\ref{example_system_for_mixed_vol}),
\begin{equation}
M(Q_{1},Q_{2})=\mbox{ Vol}_{2}(Q_{1}+Q_{2})-\mbox{ Vol}_{2}(Q_{1})-\mbox{ Vol}_{2}(Q_{2}),\label{eq:two_variable_mixvol}
\end{equation}
where,

\begin{eqnarray}
\mbox{Vol}_{2}(Q_{1}) & = & 1,\nonumber \\
\mbox{Vol}_{2}(Q_{2}) & = & \mbox{ area of parallelogram made by}\nonumber \\& & \ensuremath{\{(0,0),(1,0),(0,1),(1,1)\}}\nonumber \\
 & + & \mbox{ area of triangle made by}\nonumber \\& & \ensuremath{\{(1,1),(1,0),(1,2)\}}\nonumber \\
 & = & 1+\frac{1}{2}=\frac{3}{2},\nonumber \\
\mbox{Vol}_{2}(Q_{1}+Q_{2}) & = & \frac{13}{2}.
\end{eqnarray}

Thus, the mixed volume for this system is $4$.

The above introduced concept of the mixed volume is extremely important because the Bernstein-Khovanskii-Kushnirenko (BKK) theorem~\cite{Bernstein75,Khovanski78,Kushnirenko76}, in combination with \cite{Li96thebkk,Mau:94,MauW:96}, states that for generic coefficients, the number of isolated solutions in is exactly equal to the (stable) mixed volume of this system counting with multiplicity, and for any particular set of coefficients, it is an upper bound.

In practice, we can use a highly sophisticated implementation of an algorithm to calculate the mixed volume of a
given system called MixedVol \cite{GLW:05} which is transplanted in {\sf PHCpack} and MixedVol-2.0 which is transplanted in {\sf HOM4PS2}.

\bibliographystyle{IEEEtran}
\bibliography{bib.bib}

\end{document}